\documentclass{iau-JDSS}
\usepackage{graphicx}

\title[Cosmic rays in magnetized intracluster plasma] 
{Cosmic rays in magnetized \break intracluster plasma}

\author[L. Feretti et al.]   
{L. Feretti$^1$
A. Bonafede$^1$, G. Giovannini$^{1,2}$, F. Govoni$^3$, 
M. Murgia$^3$}

\affiliation{$^1$ Istituto di Radioastronomia INAF, Via Gobetti 101, 40129 Bologna, Italy 
\\[\affilskip]
$^2$ Dipart. Astronomia, Univ. Bologna, Via Ranzani 1, 40127 Bologna, Italy
\\[\affilskip]
$^3$ Oss. Astronomico Cagliari, Loc. Poggio dei Pini, Strada 54, 09012 Capoterra (CA), Italy
} 
\pubyear{2009}
\volume{Volume 15}  
\pagerange{1--2}
\date{?? and in revised form ??}
\jname{Highlights of Astronomy, Volume 14}
\editors{Ian F Corbett, ed.}
\begin{document}

\maketitle

\firstsection 
\section{Magnetic Fields in Clusters of Galaxies}

A breakthrough in the studies of magnetic fields in clusters of galaxies
has been reached in recent years from the analysis  of the Rotation
Measure of sources seen through the magnetized cluster medium
(Govoni \& Feretti 2004). The results obtained can be summarized as follows:
(i) magnetic fields are present in all clusters; (ii)
at the center of clusters undergoing merger activity the field
strenght is around 1 $\mu$G, whereas at the center of relaxed 
cooling core clusters the intensity is much higher ($\sim$ 10 $\mu$G); (iii) a model involving a single magnetic 
field coherence scale is not
suitable to describe the observational data, because of different
scales of field ordering and tangling.  

Assuming a magnetic field power 
spectrum: $|B_{\kappa}|^2 \propto
{\kappa}^{-n}$ (Murgia et al. 2004), the range of spatial scales is found between $30-500$ kpc
and the spectral index $n$ is in the range $2-4$ (note that in A2255, Govoni 
et al. 2006 obtain a flatter index at the 
center, and a steeper index at the periphery, likely due to 
different turbulence development).

In addition, the cluster magnetic field intensity shows a radial decline
linked to the thermal gas density n$_e$ as
$B \propto $n$^x_e$. A trend with $x$ = 1/2 is expected if the 
B field energy scales as  the thermal energy, while $x$ = 2/3 
if the B field results from a frozen-in field during the cluster collapse.
The values of $x$ derived so far are in the range
$0.5-1$. 

\section{Diffuse Radio Emission}

Magnetic fields at the $\mu$G level in the intracluster plasma are illuminated
by cosmic rays, which give rise to diffuse cluster radio emission of 
synchrotron origin. While magnetic fields are ubiquitous in clusters, 
the radio emitting electrons are currently not known to be present in all
clusters, although their presence is revealed in
several conditions  (merging and relaxed clusters),
at different cluster locations (center, periphery,  intermediate
distance), on very different scales (100 kpc to $>$Mpc), and  generally
is related to a high degree of magnetic field ordering.
Most spectacular examples of diffuse radio emission are the giant 
radio halos and relics  detected in merging clusters.
Several giant double relics, located on opposite
side with respect to the cluster center, are presently known 
(e.g. Bonafede et al. 2009, van Weeren et al. 2009). 
Radio halos of smaller size have also been detected, as
well as mini-halos and small relics in cooling core clusters.
New halos have been detected in A851, A1213, A1351, A1995, A2034 and
A2294 (Giovannini et al. 2009, also Giacintucci et al. 2009 for A1351). 
Among them, the cluster A1213 is 
remarkable because its X-ray luminosity is about 10 times  weaker than that
associated with clusters hosting radio halos (as derived so far).
 
All diffuse radio sources have in common the very steep radio spectra, implying
that the radiating particles have short lifetimes, and need to be
reaccelerated.

\section{Radio - X-ray connection in radio halos}

The properties of radio halos are linked to the properties of the host
clusters (Cassano et al. 2006, Giovannini et al. 2009), in particular:
a) the radio power of both small and giant halos correlates with the 
cluster X-ray luminosity (i.e. gas temperature and total mass);
b) the radio spectra of halos are affected by the cluster temperature, being
flatter in hotter clusters, and in hotter cluster regions 
(Orr\'u et al. 2007); 
c) in a number of well-resolved clusters, a point-to-point
spatial correlation is observed between the halo radio brightness 
 and the cluster X-ray brightness (Govoni et al. 2001).

A step forward is to check whether the last property is common to all
radio halos,
i.e. whether radio halos are generally distributed as the X-ray thermal gas.
Using a sample of clusters having good radio and X-ray data, we have
analysed the position of the radio halo with respect to that of  the X-ray gas 
distribution. The left panel of Fig. 1 shows that both giant and small 
radio halos can be significantly shifted, up to hundreds 
kpc, with respect to the centroid of the host cluster.
To highlight radio halos with the most pronounced asymmetric distribution,
we have then derived the ratio between the radio-X-ray offset and 
the halo size. From the middle and right panels of Fig.1, 
we deduce that halos can be very asymmetric with
respect to the X-ray gas distribution, and this becomes more relevant 
when halos of smaller size are considered.  
A possibility is that the asymmetry 
in the structure originates by magnetic field fluctuations as 
large as hundreds of kpc, as suggested by Vacca et
al. (2009) on the basis of magnetic field modeling.

\begin{figure}
 \includegraphics[width=14 cm, bb=50 150 580 340,clip]{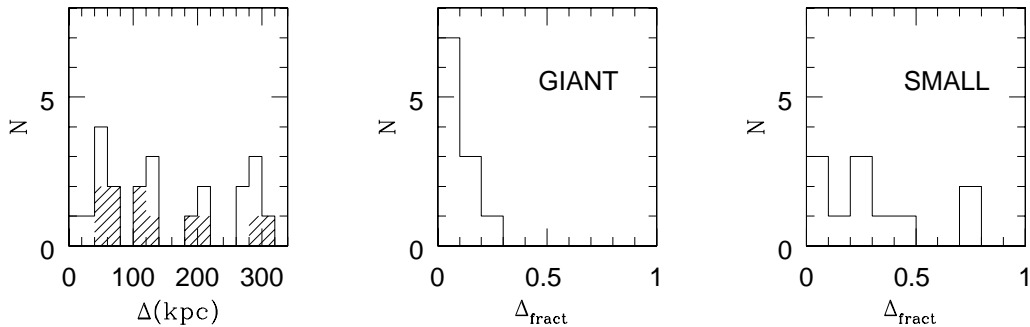}
  \caption{Left panel: Values of the offset $\Delta$ between the 
radio and X-ray centroids in kpc. The dashed area refers to giant halos.
Middle and Right panels : Fractional offset ($\Delta$/radio halo size)
for giant halos (size $\ge$ 1 Mpc)
and small halos (size $<$ 1 Mpc), respectively.
}
\end{figure}


\begin{thebibliography}{}

\bibitem[]{}
Bonafede, A., Giovannini, G., Feretti, L., Govoni, F., Murgia, M., 2009, A\&A 
494, 429

\bibitem[]{}
Cassano R., Brunetti G., Setti G.,  2006 MNRAS 369, 1577

\bibitem[]{}
Giacintucci, S., Venturi, T., Cassano, R., et al., 2009,
ApJl in press, eprint arXiv:0909.0437

\bibitem[]{}
Giovannini, G., Bonafede, A., Feretti, L.,  et al., 2009, A\&A in press,
eprint arXiv0909.0911

\bibitem[]{}
Govoni, F., En{\ss}lin, T.A.,  Feretti, L., Giovannini, G., 2001, 
A\&A  369, 441

\bibitem[]{}
Govoni, F., Feretti, L., 2004,
Int. J. Mod. Phys. D, Vol., 13,  1549

\bibitem[]{}
Govoni, F., Murgia, M., Feretti, L., et al., 2006, A\&A 460, 425

\bibitem[]{}
Murgia, M., Govoni, F., Feretti, L. , 2004, A\&A 424, 429

\bibitem[]{}
Orr\'u, E., Murgia, M., Feretti, L., et al., 2007, A\&A 467, 943

\bibitem[]{}
Vacca V., Murgia, M., Govoni, F., et al., 2009, A\&A, Submitted

\bibitem[]{}
van Weeren, R.J., et al.  2009,
A\&A in press, eprint arXiv0908.0728

\end{thebibliography}
\end{document}